\documentclass[apj]{emulateapj}
\usepackage{apjfonts}

\newcommand{\mthco}{{\rm ^{13}CO}}

\newcommand{\thco}{$^{13}$CO}
\newcommand{\co}{$^{12}$CO}

\newcommand{\av}{$A_{V}$}
\newcommand{\mav}{A_V}

\newcommand{\kms}{ { \rm km\, s^{-1} } }

\newcommand{\clfind}{{Clumpfind}}

\begin{document}
\title{The Perils of Clumpfind: The Mass Spectrum of Sub-structures in Molecular Clouds}
\shorttitle{\clfind: Sub-structures in Molecular Clouds}

\shortauthors{J. E. Pineda,  E. W. Rosolowsky, and A. A. Goodman}
\author{Jaime E. Pineda\altaffilmark{1}, Erik W. Rosolowsky\altaffilmark{2}, and Alyssa A. Goodman\altaffilmark{1}
}
\altaffiltext{1}{Harvard-Smithsonian Center for Astrophysics, 60 Garden St., Cambridge, MA 02138, USA}
\altaffiltext{2}{University of British Columbia Okanagan, 3333 University Way, Kelowna, BC V1V 1V7, Canada}
\email{jpineda@cfa.harvard.edu}
\slugcomment{Draft version 8.0, May/29/2009, JEP}

\begin{abstract}
We study the mass spectrum of sub-structures in the Perseus Molecular Cloud Complex traced by \thco~(1--0), finding that $dN/dM\propto M^{-2.4}$ for the standard \clfind\ parameters. 
This result does not agree with the classical $dN/dM\propto M^{-1.6}$. 
To understand this discrepancy we study the robustness of the mass spectrum derived using the \clfind\ algorithm.
Both 2D and 3D \clfind\ versions are tested, using 850~$\mu$m dust emission and \thco\ spectral-line observations of Perseus, respectively. 
The effect of varying threshold is not important, but varying stepsize produces a different effect for 2D and 3D cases. 
In the 2D case, where emission is relatively isolated (associated with only the densest peaks in the cloud), the mass spectrum variability is negligible compared to the mass function fit uncertainties. 
In the 3D case, however, where the \thco\ emission traces the bulk of the molecular cloud, the number of clumps and the derived mass spectrum are highly correlated with the stepsize used. 
The distinction between ``2D" and ``3D" here is more importantly also a distinction between ``sparse" and ``crowded" emission.  
In any ``crowded" case, \clfind\ should not be used \emph{blindly} to derive mass functions.  
\clfind's output in the ``crowded" case can still offer a statistical description of emission useful in inter-comparisons, but the clump-list should not be treated as a robust region decomposition suitable to generate a physically-meaningful mass function.
We conclude that the \thco\ mass spectrum depends on the observations resolution, due to the hierarchical structure of MC. 
\end{abstract}

\section{Introduction}
Molecular clouds (MCs) have usually been studied using \co\ and \thco~(1--0) transition line maps, because they trace low-density material. 
When these observations are used the emission comes from the whole MC, and the emission is ``crowded''.
They also provide information about the velocity structure of the cloud, and we refer them as 3D data.
Thanks to the new generation of bolometers, large scale dust emission maps of entire MCs are now possible \citep{Motte:Oph-IMF,Hatchell:SCUBA,clump-2d,Kirk_2006-Perseus,Enoch:Perseus}.  
But, due to the observing technique much of the emission on large-scales is removed, obtaining a map with ``sparse" emission.
These dust emission maps are mostly used to find the densest objects in a MC: dense cores.
However, these data do not provide velocity information to asses if a core is bound or not; and we refer these data as 2D.
Extinction maps provide another tool to study MCs and dense cores using 2D data \citep{Cambresy:1999,Lombardi_2006-Pipe_NICER}. 
These maps give an estimate of the total column density in a region hence capturing the large-scale structures in MCs, and therefore the map is crowded. 
However, thanks to some post-processing techniques the extended emission can be removed, finally obtaining a map with just sparse emission \cite{Pipe:cores}.
Finally, by observing molecular lines with higher critical density or with an interferometer most of the large-scale structure is not traced, obtaining a 3D data set but with sparse emission.

The structure of MCs has been studied using a variety of decomposition algorithms on  \co\ and/or \thco~(1--0) emission maps  \citep[e.g.][]{gaussclumps,gaussclump-dndm,clumpfind,Williams:Rosette}. 
Algorithms that decompose the MC typically take all the emission (above some threshold) and split it into clumps, which can later be easily used to calculate a mass function. 
In such studies, it has been shown that the mass function of clumps follows a power-law with $dN/dM \propto M^{-1.6\pm 0.2}$ \citep{Blitz:ppIII}. 

One of the most widely used cloud decomposition algorithms is \clfind\ \citep{clumpfind}, since it is readily available and has only two user-controlled parameters. It was designed to study a whole MC using 3D molecular line data in a systematic fashion \citep{clumpfind,Williams:Rosette}, and 
the data historically had coarse angular resolution allowing the study of only the largest structures in the cloud.
\clfind\ has also been modified to handle 2D data with sparse emission, and successfully applied to study the core mass function \citep[CMF, e.g.][]{clump-2d,Kirk_2006-Perseus,Pipe:cores,Reid:summary,Reid:M17,reid:NGC7538}, and to compare that CMF to the Initial Mass Function (IMF) of stars which appears as an almost invariant power-law \citep{Salpeter:IMF,Muench:IMF,Kroupa:IMF}: $dN/dM\sim M^{-2.35}$ for $M>0.6~M_\odot$.
In just a few cases molecular line data from a higher density tracer has been used to study dense cores \citep{Ikeda-Orion:2007,Walsh:NGC1333}, adding velocity information to the mostly sparse emission.

In this work, we study the robustness of the mass spectrum derived using \clfind\ in crowded (\thco~(1--0)) and sparse emission (SCUBA 850$\mu$m). 
By using the \thco~(1--0) and SCUBA 850$\mu$m data collected by the COMPLETE team in the Perseus Molecular Cloud Complex \citep{COMPLETE-I,Kirk_2006-Perseus}\footnote{All of the data from the COMPLETE (COordinated Molecular Probe Line Extinction Thermal Emission) Survey are available on-line at \url{http://www.cfa.harvard.edu/COMPLETE}.} we are able to study the algorithm in both its 3D and 2D versions on real data sets (with overlapping sky coverage) with high resolution and sensitivity. 

\section{Data}\label{data}
We use the \thco~(1-0) molecular line map obtained by the COMPLETE Survey \citep{COMPLETE-I} using the SEQUOIA 32-element focal plane array at the FCRAO telescope. Observations were carried out using the on-the-fly technique.
The data cube covers an area of $\sim$6.25$\degr\times$3$\degr$ with a $46\arcsec$ 
beam on a $23\arcsec$ grid, and it is presented in $T_A^*$ scale.

The map is beam sampled and Hanning smoothed in velocity.
The final pixel size is $46\arcsec$ and the velocity resolution is $\Delta v=0.066~\kms$.
The median Root Mean Square (RMS) noise in the map is $0.1~{\rm K}$ in $T_A^*$, and all positions with ${\rm RMS}>0.3~{\rm K}$ are removed from the map. In addition, a noise-added \thco\ cube (with RMS=0.2~K) is also used.

We also use the 850~$\mu$m map obtained with SCUBA on the JCMT \citep{Kirk_2006-Perseus}. 
The pixel scale is 6\arcsec, while the effective beam is 19.9\arcsec. 
The mean RMS in the map is $\sim$0.06~Jy\,beam$^{-1}$. The map coverage is smaller than the \thco\ data, but it covers the densest regions in the cloud, where the dense cores identified by SCUBA lie \citep{Kirk_2006-Perseus,Hatchell:SCUBA}. 
It is important to note that in the SCUBA map any structure larger than $\sim$2$\arcmin$ is removed during the data reduction process (in addition to the observational problems of detecting extended structure with bolometers), making the map mostly devoid of extended emission. Hence the SCUBA map is substantially different from the \thco\ data, because the later traces the more extended material.
Small areas near the SCUBA map's edge are removed to avoid some image artifacts.

\section{Cloud Structure}
\subsection{Clump Identification}
\clfind\ needs only two parameters (threshold and stepsize) to decompose the emission onto a set of clumps. The threshold parameter sets the minimum emission required to be included in the decomposition, while the stepsize defines how finely separated the iso-surfaces (or iso-contours) are drawn by the algorithm in order to check for structures.
In other words, threshold sets the number of pixels included in the decomposition, and stepsize sets the contrast needed between two features to be identified as different objects.
\cite{clumpfind} suggest using a threshold and stepsize value of 2--$\sigma$, where $\sigma$ is the noise in the data. 
\clfind\ assigns all the emission above the given threshold into clumps, and it can be applied to both 2D and 3D data sets.

For this analysis, different values for both parameters are used to test the robustness of the results derived using \clfind. 
In the 3D case (\thco), the threshold is set to 3--, 5-- and 7--$\sigma$ for the original data and 5--$\sigma$ for the noise-added one; 
while the stepsize is varied between 3-- and 20--$\sigma$ with an spacing of 0.5--$\sigma$. 
In the 2D case (SCUBA), the threshold is set to 
3--, 5--, and 7--$\sigma$; 
and the stepsize is varied between 
2-- and 16.5--$\sigma$, with a spacing of 0.5--$\sigma$.

Some stepsize values seem unusually large, but given the improvement on the data available larger stepsizes are required to identify the largest structures in MC \citep[see][]{Rathborne_2009-GRS_clumps}

\subsection{Mass Estimate}
We adopt the conversion between \thco\ integrated intensity, $W(\mthco)$, and extinction, \av, derived by \cite{pineda:abundance:pers}:
\begin{equation}
\mav = 0.350\,W(\mthco)~.
\end{equation}
This conversion is derived for Perseus using the COMPLETE extinction map and FCRAO data  (assuming a main-beam efficiency of $0.49$).
To convert from visual extinction to column density, we assume 
that  the ratio between $N(\rm{H})$ and $E(B-V)$ is $5.8\times 10^{21}~\rm{cm^{-2}\,mag^{-1}}$ 
\citep{bohlin:1978}, and  $R_V=3.1$.

For the dust continuum emission, we assume it is optically thin,
\begin{equation}
M_{850} = 0.48\, S_{850} \left(\frac{\kappa_{850}}{0.02~{\rm cm^2\,gr^{-1}}} \right)^{-1}~M_\odot~,
\end{equation}
where $S_{850}$ is the flux at $850~{\rm \mu m}$, $\kappa_{850}$ is the opacity at 850~$\mu$m, and we assume a dust opacity of 0.02~${\rm cm^2\,gr^{-1}}$ \citep{OH:94}, dust temperature of $T_d=15$~K and a distance to Perseus of 250~pc. These adopted values are the same used by \cite{Kirk_2006-Perseus}.

\subsection{Completeness limit}
\begin{figure*}
\plotone{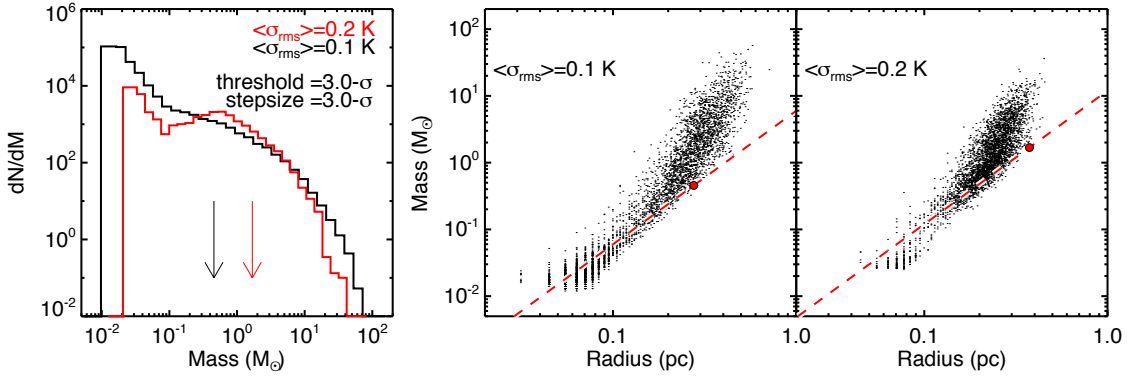}
\caption{Comparison of clump mass functions for original and noise-added \thco\ data. 
\emph{Left} panel shows the change in mass function of two \clfind\ runs using the same threshold and stepsize (in $\sigma$ units), but with different noise levels. 
Arrow shows the corresponding completeness limit.
\emph{Middle} and \emph{right} panels show the mass and radius for each identified clump. Red dashed line shows the sensitivity curve, and the red filled circle shows the sensitivity limit estimated for each \clfind\ run.
The possible mass function turnover is close to where the identified clumps cross the sensitivity curve.
\label{fig:comp}}
\end{figure*}

For the \thco\ data, the completeness limit is estimated by comparing the derived mass and radius of each clump and a sensitivity curve. 
The sensitivity curve is estimated as the smallest realistic clump that can be found by \clfind\ given a radius (red dash line in Figure~\ref{fig:comp}). 
The completeness limit is then estimated as the largest mass below this sensitivity curve (for each \clfind\ run), i.e. the mass where data and red line merge in Figure~\ref{fig:comp} and shown with red filled circle. 
Here, a minimum size of 3 velocity channels ($\Delta v$) is assumed, and given that the brightness in each pixels must be larger than the threshold: 
\begin{equation}
M_{min}(R)=0.091\left(\frac{3 \Delta v}{{\rm km\,s^{-1}}}\right)\left(\frac{\pi R^2}{46\arcsec^2}\right)\left(\frac{\xi}{K}\right)M_\odot,
\end{equation}
where $\xi$ is the threshold. 
In Figure~\ref{fig:comp} two \clfind\ runs are shown: original and noise-added \thco.
In both cases the possible change in slope of the mass function happens close to the completeness limit shown by the arrow.

For simplicity, we use a single completeness limit for each dataset, a value larger than the completeness limit estimated for any of the individual \clfind\ runs: 
4 and 3~$M_\odot$ for the original and noise-added \thco, respectively.
The completeness limit for the objects identified in the SCUBA map is estimated by \cite{Kirk_2006-Perseus} as 0.6~$M_\odot$, not as the mass where the mass function changes, but as the object that would be missed given the typical size of the cores found.

\section{Results}

\begin{figure*}
\plottwo{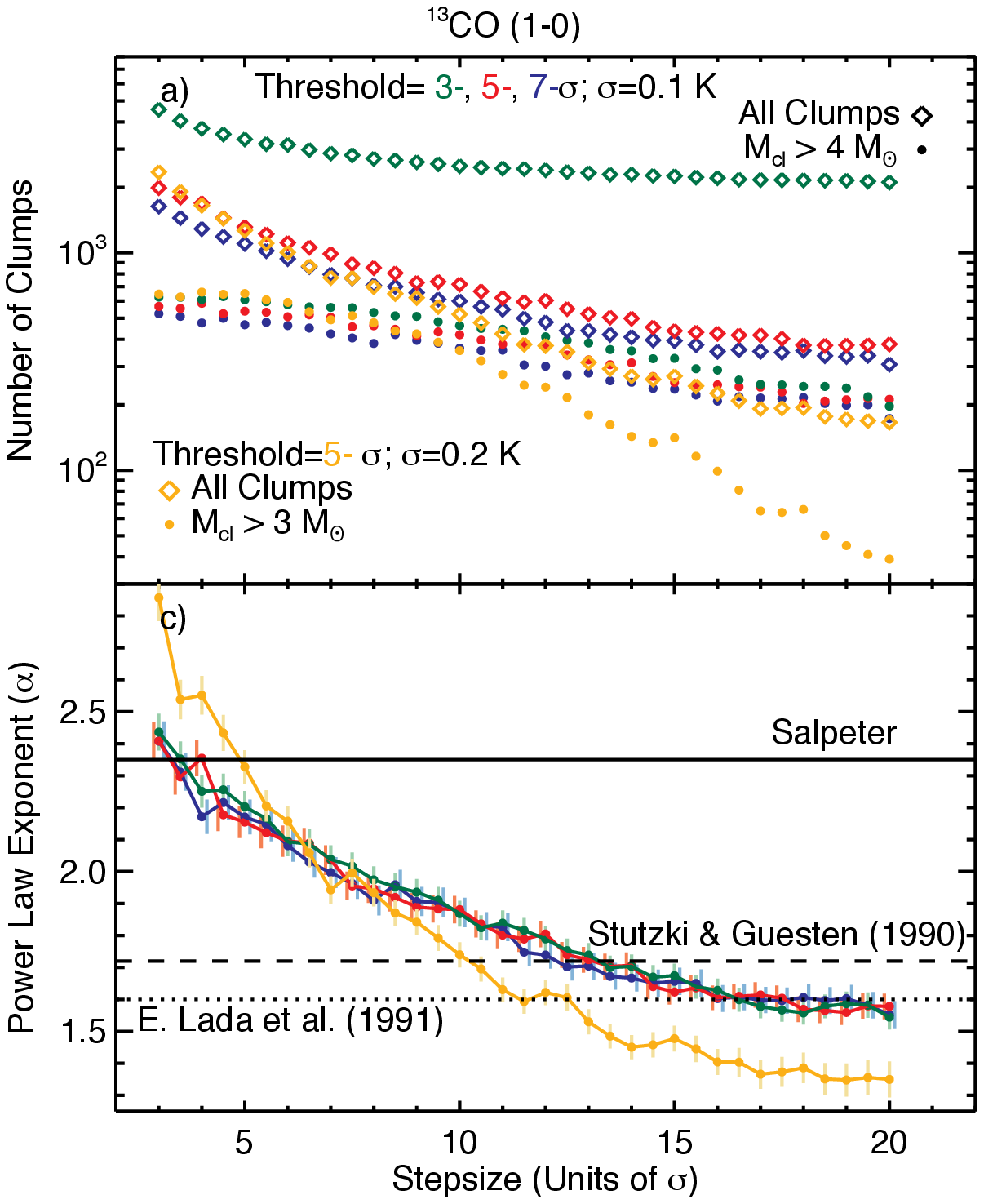}{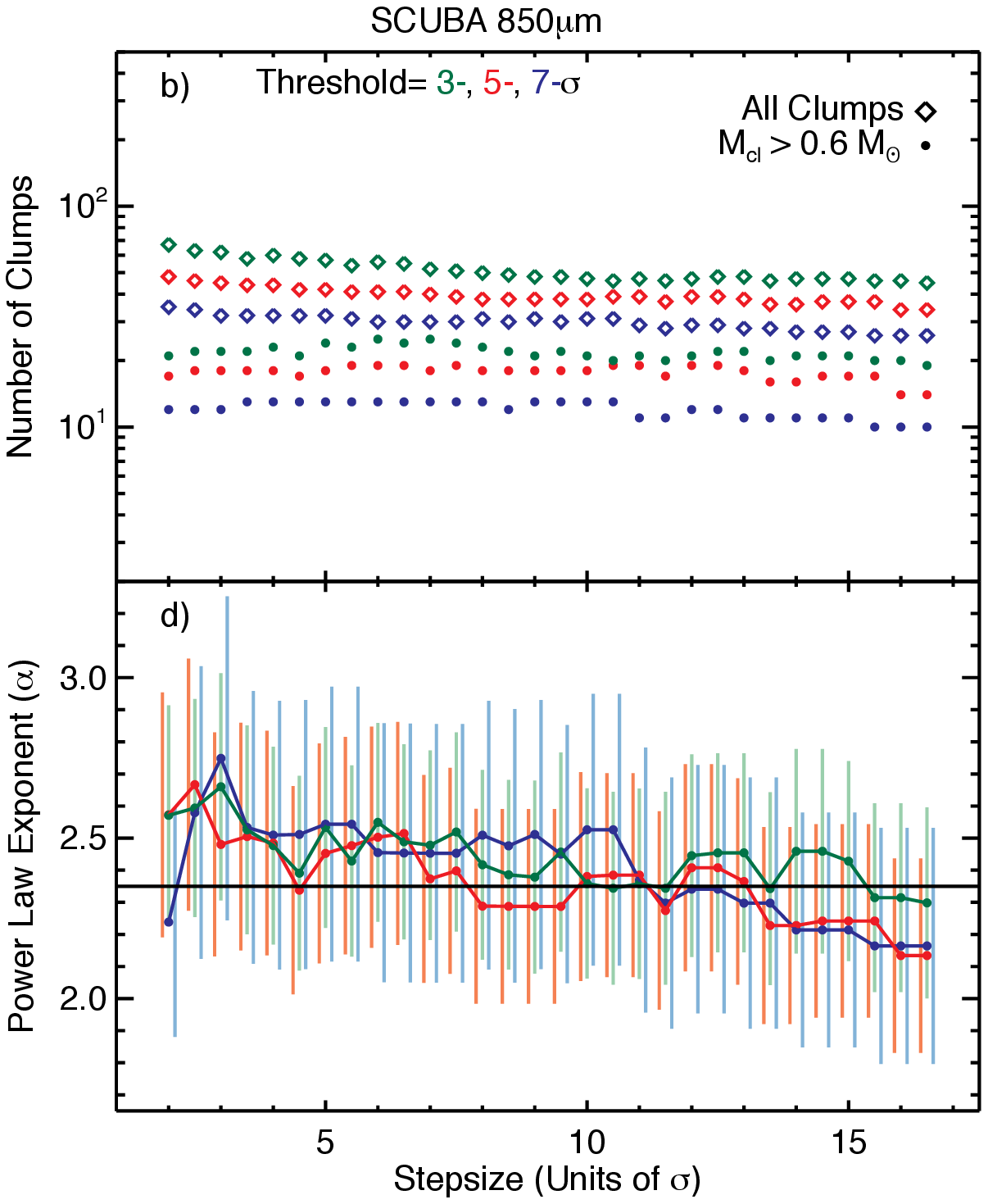}
\caption{
Summary of all \clfind\ runs as a function of stepsize. 
Color represent different Thresholds: blue, red, and green for 3--, 5--, and 7--$\sigma$, respectively; we also show in orange results with Threshold of 5--$\sigma$ for \thco\ data with added noise. \emph{Left} and \emph{right} columns show results for \thco\ and SCUBA data, respectively.
Panels \textbf{a} and \textbf{b} show the number of clumps under a given category per model.
Total number of clumps found, and total number of clumps with mass larger than the completeness limit are shown in open diamonds and filled circles, respectively.
Panels \textbf{c} and \textbf{d} show the exponent of the fitted mass spectrum of clumps above the completeness limit, $dN/dM\propto M^{-\alpha}$, with error bars estimated from equation~\ref{eq-sigma}. 
Horizontal black lines show some fiducial exponents for comparison. 
Average noise in: \thco, \thco\ with added noise and SCUBA data is 0.1~K, 0.2~K and 0.06~Jy\,beam$^{-1}$, respectively.
Completeness limit is estimated to be $4~M_\odot$, $3~M_\odot$ and $0.6~M_\odot$ for \thco, \thco\ with added noise and SCUBA data. 
Panel \textbf{c} also shows that for different noise level in the data, if a threshold of $\sim$2~K (20-- and 10--$\sigma$ for original and noise-added data, respectively) is used, then the fitted power-law exponents are closer to previous works.
\label{fig:summary}}
\end{figure*}

The simplest comparison between different \clfind\ runs is how many clumps are defined. In panels \textbf{a} and \textbf{b} of Figure~\ref{fig:summary} we show that the total number of clumps identified in each run (filled circles) decreases when increasing the threshold or stepsize, in either 2D or 3D. This decrease is not a surprise, because with a higher threshold there are fewer pixels available, and therefore a smaller volume to define clumps; in the case of the stepsize, a larger stepsize can miss some real structure, but also small stepsize can identify spurious clumps from structure due to noise (i.e. split a single clump into two or more because the noise creates fake structure above the stepsize level).
However, \clfind\ runs with thresholds of 3--$\sigma$ can identify twice as many clumps as runs with higher thresholds and the same stepsize (see panels {\bf a} and {\bf b} in Figure~\ref{fig:summary}), while runs with 5-- and 7--$\sigma$ thresholds follow a similar curve (with the runs of lower threshold still finding more clumps as expected) for a given data set.
It is important to note that objects identified in \thco\ are not necessarily bound.

Despite the difference in the total number of clumps, the number of clumps above the completeness limit (shown as filled circles in panels \textbf{a} and \textbf{b} of Figure~\ref{fig:summary}) is comparable between different \clfind\ runs. Not only between different thresholds, but also when changing stepsize. 
However, the number of clumps above the completeness limit is usually less than half the total number of clumps, and therefore, most of the identified clumps are not even considered in mass function analysis.

The differential mass function, $dN_{cl}/dM\approx\Delta N_{cl}/\Delta M$, is usually approximated by a power-law, 
$dN_{cl}/dM\propto M^{-\alpha}$. However, if the data are binned, then variations in the fitted power-law exponent are generated by changing the bin width and shifting the bins \citep{Rosolowsky:dNdM}; and when analyzing the cumulative function special care must be taken to avoid the undesired effects of truncation \citep{munoz:cores,Li:cores}. To avoid both problems we perform a fit of the differential mass function, \emph{but without binning the data}, by using the Maximum Likelihood Estimate \citep[MLE, see][]{Clauset_2007-power_law_fit}.

We fit the following function,
\begin{equation}\label{eq:dndm}
\frac{dN}{dM} = N_{cl} \frac{\alpha -1 }{M_{min}} \left( \frac{M}{M_{min}} \right)^{-\alpha}~,
\end{equation}
where $M_{min}$ is the minimum mass of the sample to be used in the fitting, $N_{cl}$ is the number of clumps more massive than $M_{min}$, and $\alpha$ is the power-law exponent of the distribution. Using MLE, the exponent is estimated by
\begin{equation}
\alpha = 1+ N_{cl}\left[ \sum_{i=1}^{N_{cl}} \ln{ \frac{M_i}{M_{min}} } \right]^{-1}~,
\end{equation}
and the standard error on $\alpha$ is approximated by 
\begin{equation}\label{eq-sigma}
\sigma_\alpha=\frac{\alpha-1}{\sqrt{N_{cl}}}~.
\end{equation}
This estimate can be regarded as a lower limit in the uncertainty, because it does not take into account uncertainties in the mass measurements. 
For this work we use $M_{min}$ equals to the completeness limit.

The exponent, $\alpha$, estimated for every \clfind\ run is shown as a function of stepsize in panels \textbf{c} and \textbf{d} in Figure~\ref{fig:summary} for \thco\ and SCUBA, respectively. 
The derived values for the clump mass spectrum from \cite{Lada-Lynds,gaussclumps} (using different methods and in different regions) and Salpeter's exponent (for the IMF) are also shown for comparison. 
For a standard stepsize of 3--$\sigma$, the \thco\ clump mass spectrum is similar to the IMF, and steeper than the values derived by previous works.

An interesting result is that the clump mass spectrum agrees (within the uncertainties of the fit) for different threshold values used if the same stepsize is used. 
However, most important is the fact that the estimated power-law exponent, $\alpha$, is correlated with the stepsize. 
This variation in $\alpha$ can be as high as 40\%, and the correlation appears in both versions of \clfind: 3D and 2D. 
For 2D \clfind, we find that this correlation is negligible compared with the uncertainties associated with the fitted power-law exponent. 

\section{Discussion}
\begin{figure*}
\plotone{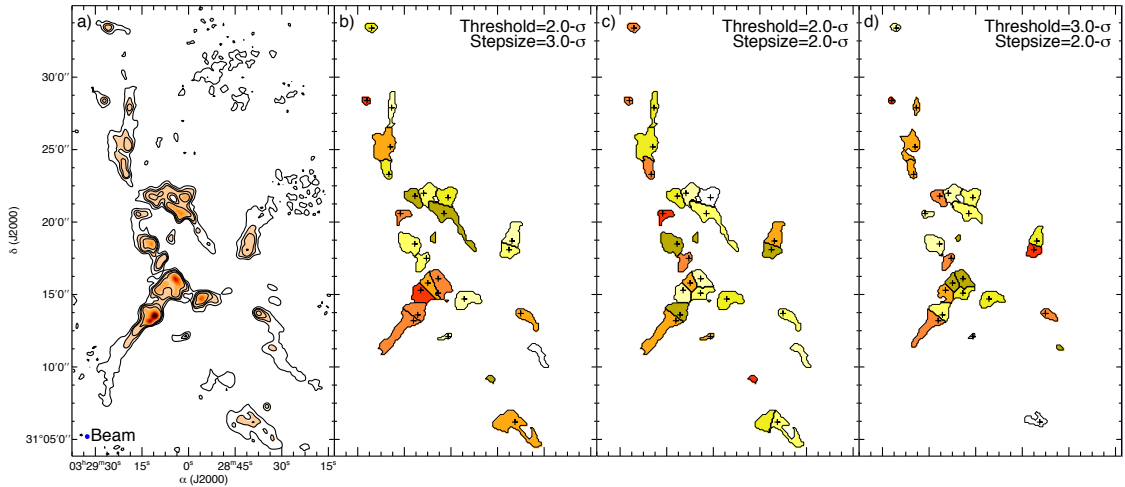}
\caption{Comparison of different \clfind\ runs on the NGC1333 region. 
Panels {\bf b}, {\bf c} and {\bf d} show clumps found in three different \clfind\ runs, and crosses mark the position of cores found by \cite{Kirk_2006-Perseus}. 
Panel {\bf a} shows the dust emission map in the NGC1333 region, with the overlaid contours at 1--, 3--, 5--, and 10--$\sigma$ level; in addition the beam size (19.9\arcsec) is shown in bottom left corner. 
Small changes in the parameters used generate small (but important) changes the catalogue obtained.
\label{fig-maps}}
\end{figure*}

From Figure~\ref{fig:summary} we can clearly see that the power-law exponent fitted to the decomposition done by \clfind\ of the \thco\ data is strongly correlated with the stepsize, and therefore not unique. 
In fact, our results show that \clfind\ is not very useful to identify small structures within a map, unless they are isolated.
The reason for this correlation between fitted power-law and stepsize is that for a small stepsize less contrast is required to identify the structure, generating more but smaller objects 
and therefore having a steeper mass distribution; this effect is more important in crowded regions (see Figure~\ref{fig-maps}). 
Despite the fact that the previous conclusion seems obvious, the amount of variation in the fitted exponent has typically been deemed negligible.
An independent analysis carried out by \cite{clfind:simulation} on numerical simulations also found different results from \clfind\ when investigating the effect of the data resolution on the \clfind\ analysis. \cite{clfind:simulation} ``observe''  a numerical simulation using different spatial resolutions for the final ``data'', and then run \clfind\ on them.
They show that for the same region, \clfind\ identifies a different number of clumps and their derived properties are variable when using different spatial resolution. 


In the 2D case, we notice that the exponent does not vary significantly with stepsize. 
In addition, the exponent fluctuation is almost negligible when compared with the associated uncertainties. 
However, Figure~\ref{fig-maps} shows an example of how different the \clfind\ decomposition is for three different input parameters. 
By comparing different panels in Figure~\ref{fig-maps} we see that  some structures appear or are split under different parameters. 
These subtle differences suggest that to create a reliable catalogue manual check is needed to ensure meaningful structures.
Moreover, \cite{Kainulainen_2009-fidelity_cmf} recently showed, using the Pipe MC extinction data, that the CMF can not be recovered in crowded cases.

\clfind\ is also run on the noise-added \thco\ data with 5--$\sigma$ threshold. 
The fitted power-law exponents for noise-added \thco\ structures are similar to those derived for the original \thco\ data \emph{only} for large stepsizes ($\sim2$~K).
Also, only for large stepsizes the fitted power-law exponent is close to results from previous studies of the structure in MCs. 
But, this should not be a surprise, since \clfind\ will assign the \thco\ extended emission into several clumps, and by adding noise the boundaries of these clumps are changed. 
This generates more less-massive clumps and also changes the slope of the power-law. 
However, there must be a point where \clfind\ identifies the largest structures in the cloud and the exponent should not change much for larger stepsize. 
We estimate that this effect must be less dramatic when the emission is sparse (e.g.  SCUBA map, interferometer data or higher density tracer), because there is less room to change the boundaries and masses of the objects. 
Also, a different structure identification techniques, dendrogram \citep{dendrogram,Nature:3d-pdf}, that allows for hierarchical structure is already available and could be used to derive mass function of bound structures or any specific structure under consideration.

\section{Summary}
The \thco\ and 850~$\mu$m maps of Perseus of the COMPLETE Survey are used to study the 2D and 3D versions of the \clfind\ algorithm, respectively.

The total number of identified structures is highly correlated with the parameters used (threshold and/or stepsize). 
Decompositions run with a smaller threshold and stepsize produce more objects. 

We use a new method to estimate the completeness limit for a sample of clumps.
The mass spectrum of the identified structures, $dN/dM$, is fitted with a power-law above the completeness limit. 
For the standard \clfind\ parameters, the mass function exponent is closer to Salpeter than to the classical result from \cite{Blitz:ppIII}.
Despite the small variation in the number of objects above the completeness limit, the fitted power-law exponent for \thco\ structure is a strong function of the stepsize, while it is independent of the threshold used. 
The power-law exponent of SCUBA objects is also correlated with stepsize, but this effect is negligible compared to the associated uncertainties from the fitting.
The \thco\ power-law exponent variation shows that the cloud structure changes as we approach smaller scales, and that \clfind\ is still a useful tool to study the structure of a Molecular Cloud or the difference between two regions. 
However, this also means that it is not possible to derive a \emph{single} mass function describing the sub-structure in molecular clouds when using a non-hierarchical decomposition. 
Most likely, a better way to study the structure in molecular clouds is by using some identification scheme that takes into account the hierarchical nature of these regions (e.g. dendrograms). In such case, mass distribution functions of the bound material could be used as an observable.



\acknowledgments
We thank Jonathan B. Foster, and Jens Kauffmann for helpful discussions.
We thank the referee for useful comments that helped to improve the paper.
JEP is supported by the National Science Foundation through grant 
\#AF002 from the Association of Universities for Research in 
Astronomy, Inc., under NSF cooperative agreement AST-9613615 and by 
Fundaci\'on Andes under project No. C-13442. 
This material is based 
upon work supported by the National Science Foundation under Grant 
No. AST-0407172 to AAG. 
EWR's work is supported by an NSF Astronomy and Astrophysics Postdoctoral Fellowship (AST-
0502605) and a Discovery Grant from NSERC of Canada.


\end{document}